\documentclass[superscriptaddress,prl,twocolumn,nofootinbib,showpacs,preprintnumbers]{revtex4}

\usepackage{amsfonts}
\usepackage{mathrsfs}
\usepackage{amssymb}
\usepackage{amsmath}
\usepackage{graphicx,epsfig}

\def\bwt{\begin{widetext}}

\def\ewt{\end{widetext}}

\def\be{\begin{equation}}

\def\ee{\end{equation}}

\def\bea{\begin{eqnarray}}

\def\eea{\end{eqnarray}}

\def\bean{\begin{eqnarray*}}

\def\eean{\end{eqnarray*}}

\def\bary{\begin{array}}

\def\eary{\end{array}}

\def\bit{\begin{itemize}}

\def\eit{\end{itemize}}

\def\su5u1{SU(5) \times U(1)}

\def\fsu5u1{SU(5) \times U(1)'}

\def\so10{SO(10)}

\def\sq20{SO(10) \times SO(10)}

\usepackage{amssymb}

\begin{document}

\title{ Fast Proton Decay}

\author{Tianjun Li}

\affiliation{George P. and Cynthia W. Mitchell Institute for
Fundamental Physics, Texas A$\&$M University, College Station, TX
77843, USA }

\affiliation{Key Laboratory of Frontiers in Theoretical Physics,
      Institute of Theoretical Physics, Chinese Academy of Sciences,
Beijing 100190, P. R. China }

\author{Dimitri V. Nanopoulos}

\affiliation{George P. and Cynthia W. Mitchell Institute for
Fundamental Physics,
 Texas A$\&$M University, College Station, TX 77843, USA }

\affiliation{Astroparticle Physics Group,
Houston Advanced Research Center (HARC),
Mitchell Campus, Woodlands, TX 77381, USA}

\affiliation{Academy of Athens, Division of Natural Sciences,
 28 Panepistimiou Avenue, Athens 10679, Greece }

\author{Joel W. Walker}

\affiliation{Department of Physics, Sam Houston State University,
Huntsville, TX 77341, USA }



\begin{abstract}

We consider proton decay in the testable flipped
$SU(5)\times U(1)_X$ models with TeV-scale 
vector-like particles which can be realized in free 
fermionic string constructions
and F-theory model building.
We significantly improve upon the determination of light threshold
effects from prior studies, and perform
a fresh calculation of the second loop for the
process $p\rightarrow e^+\pi^0$ from the heavy gauge boson exchange.
The cumulative result is comparatively fast
proton decay, with a majority of the most plausible
parameter space within reach of the future
Hyper-Kamiokande and DUSEL experiments.  
Because the TeV-scale vector-like particles can be
produced at the LHC, we predict a strong correlation
between the most exciting particle physics
experiments of the coming decade.

\end{abstract}


\preprint{ACT-10-09, MIFP-09-39}

\maketitle


{\bf Introduction~--}~Supersymmetry naturally solves 
 the gauge hierarchy problem of the Standard Model (SM).
Especially, in the supersymmetric SM, the three gauge 
couplings for  $SU(3)_C$, $SU(2)_L$ and $U(1)_Y$
are unified at about $2\times 10^{16}$ GeV~\cite{Langacker:1991an}. 
This strongly indicates that there may exist Grand Unified 
Theories (GUTs) at the unfication scale. Interestingly, GUTs
 give us a simple understanding of the quantum numbers for the 
SM fermions. One of the major predictions of GUTs is that the proton 
becomes destablized due to the quark and lepton unification. 
Pairs of quarks may transform into a lepton and an anti-quark
via dimension six operators from 
the exchange of heavy gauge bosons, and thus
the proton may decay into a lepton plus meson final state.  
Because the masses of heavy gauge bosons are near to the GUT scale, 
such processes are expected to be very rare. Indeed, 
proton decay has not yet been seen in the expansive
Super-Kamiokande experiment, which places a 
lower bound on such partial lifetime around
$6-8\times 10^{33}$ years~\cite{:2009gd}.

In the standard supersymmetric 
$SU(5)$ models~\cite{Georgi:1974sy, Dimopoulos:1981zb}, there exists 
the initial problem of Higgs doublet-triplet splitting, and the 
additional threat of proton decay via dimension five operators 
from exchange of the colored Higgsino (supersymmetric partners 
of the colored triplet Higgs fields)~\cite{Murayama:2001ur}.
Interestingly, when we embed the supersymmetric 
$SU(5)$ models into the M-theory model 
building~\cite{Acharya:2001gy, Witten:2001bf} 
or F-theory model building~\cite{Beasley:2008dc, Donagi:2008ca}, 
we can naturally solve the 
doublet-triplet splitting problem and the dimesnion
five proton decay problem. Moreover, in the flipped
$SU(5)\times U(1)_X$ models~\cite{smbarr, dimitri, AEHN-0},
these difficulties are solved elegantly 
due to the missing partner mechanism~\cite{AEHN-0}.
We thus only need consider dimension six proton decay.
This initial salvation from the dimension five proton decay
has sometimes turned subsequently to 
frustration~\cite{Ellis:1995at, Ellis:2002vk}
that large portions of the parameter space
in the minimal flipped $SU(5)\times U(1)_X$ model 
predict a lifetime so long as to be unobservable 
by even hypothetical proposals for future experiments.

In this paper, we consider the testable 
flipped $SU(5)\times U(1)_X$
models with TeV-scale vector-like particles~\cite{Jiang:2006hf}.
Such models can be realized within free 
fermionic string constructions~\cite{Lopez:1992kg} 
and also F-theory model building~\cite{Beasley:2008dc,
Donagi:2008ca, Jiang:2008yf}. 
Interestingly, we can solve the little hiearchy problem 
between the string scale and the GUT scale in the free 
fermionic string models~\cite{Jiang:2006hf}, and we can 
explain the decoupling scenario in F-theory models~\cite{Jiang:2008yf}. 
We undertake a highly detailed calculation of proton decay
in the dimension six $p\rightarrow e^+\pi^0$ channel,
significantly improving upon the determination light threshold
effects from prior studies, performing a fresh evaluation of the second loop,
and correcting a subtle computational inconsistency
from earlier work~\cite{Ellis:1995at, Ellis:2002vk}.
The cumulative result is a signficantly more rapid prediction 
for proton decay, with a majority of the most plausible
parameter space within reach of the future
Hyper-Kamiokande~\cite{Nakamura:2003hk} and
Deep Underground Science and Engineering 
Laboratory (DUSEL)~\cite{DUSEL} experiments.
We emphasize that the TeV-scale vector-like particles
 under consideration are accessible to the Large Hadron Collider (LHC), 
presenting a strong correlation between
that important experiment and the ongoing search for proton decay.
We also realize a dramatic shortening of the proton lifetime
in the minimal flipped $SU(5)\times U(1)_X$ model,
making detection within that scenario also quite feasible,
barring action of very large threshold corrections near the GUT scale.
Full details of our calculation will be presented 
in a subsquent report~\cite{LNW-LP}.

{\bf Flipped $SU(5)\times U(1)_X$ Models~--}~We first 
briefly review the minimal flipped
$SU(5)\times U(1)_X$ model~\cite{smbarr, dimitri, AEHN-0}. 
There are three families of SM fermions 
whose quantum numbers under $SU(5)\times U(1)_{X}$ are
\bea
F_i={\mathbf{(10, 1)}},~ {\bar f}_i={\mathbf{(\bar 5, -3)}},~
{\bar l}_i={\mathbf{(1, 5)}},
\label{smfermions}
\eea
where $i=1, 2, 3$. 

To break the GUT and electroweak gauge symmetries, we 
introduce two pairs of Higgs fields
\bea
H={\mathbf{(10, 1)}},~{\overline{H}}={\mathbf{({\overline{10}}, -1)}},
~h={\mathbf{(5, -2)}},~{\overline h}={\mathbf{({\bar {5}}, 2)}},
\label{Higgse1}
\eea
where particle assignments of the Higgs fields are 
\bea
H=(Q_H, D_H^c, N_H^c)~,~
{\overline{H}}= ({\overline{Q}}_{\overline{H}}, {\overline{D}}^c_{\overline{H}}, 
{\overline {N}}^c_{\overline H})~,~\,
\label{Higgse2} \\
h=(D_h, D_h, D_h, H_d)~,~
{\overline h}=({\overline {D}}_{\overline h}, {\overline {D}}_{\overline h},
{\overline {D}}_{\overline h}, H_u)~,~\,
\label{Higgse3}
\eea
where $H_d$ and $H_u$ are one pair of Higgs doublets in the supersymmetric SM.
We also add a SM singlet field $\Phi$.

To break the $SU(5)\times U(1)_{X}$ gauge symmetry,
we introduce the following Higgs superpotential 
\bea
{\it W}~=~\lambda_1 H H h + \lambda_2 {\overline H} {\overline H} {\overline
h} + \Phi ({\overline H} H-M_{\rm H}^2). 
\label{spgut} 
\eea 
There is only
one F-flat and D-flat direction, which can always be rotated into oriention with 
$N^c_H$ and ${\overline {N}}^c_{\overline H}$, yielding 
$<N^c_H>=<{\overline {N}}^c_{\overline H}>=M_{\rm H}$. In addition, the
superfields $H$ and ${\overline H}$ are absorbed, acquiring large masses via
the supersymmetric Higgs mechanism, except for $D_H^c$ and 
${\overline {D}}^c_{\overline H}$. The superpotential terms 
$ \lambda_1 H H h$ and
$ \lambda_2 {\overline H} {\overline H} {\overline h}$ couple the $D_H^c$ and
${\overline {D}}^c_{\overline H}$ with the $D_h$ and ${\overline {D}}_{\overline h}$,
respectively, to form heavy eigenstates with masses
$2 \lambda_1 <N_H^c>$ and $2 \lambda_2 <{\overline {N}}^c_{\overline H}>$. So then, we
naturally achieve doublet-triplet splitting due to the missing
partner mechanism~\cite{AEHN-0}. Because
the triplets in $h$ and ${\overline h}$ only have
small mixing through the $\mu h {\overline h}$ term with
$\mu$ around the TeV scale, we also solve the dimension five
proton decay problem from the colored Higgsino exchange.

In flipped $SU(5)\times U(1)_X$ models, the 
$SU(3)_C\times SU(2)_L$ gauge couplings are first joined 
at the scale $M_{23}$, and the $SU(5)$ and $U(1)_X$ gauge
couplings are subsequently unified at the higher scale $M_U$.
To separate the $M_{23}$ and $M_U$ scales
and obtain true string-scale gauge coupling unification in 
free fermionic string models~\cite{Jiang:2006hf} or
the decoupling scenario in F-theory models~\cite{Jiang:2008yf},
we introduce vector-like particles which form complete
flipped $SU(5)\times U(1)_X$ multiplets.
In order to avoid the Landau pole
problem for the strong coupling constant, we can only introduce the
following two sets of vector-like particles around the TeV 
scale~\cite{Jiang:2006hf}
\begin{eqnarray}
&& Z1:  XF ={\mathbf{(10, 1)}}~,~
{\overline{XF}}={\mathbf{({\overline{10}}, -1)}}~;~\\
&& Z2: XF~,~{\overline{XF}}~,~Xl={\mathbf{(1, -5)}}~,~
{\overline{Xl}}={\mathbf{(1, 5)}}
~.~\,
\end{eqnarray}
For notational simplicity, we define the flipped
$SU(5)\times U(1)_X$ models with $Z1$ and
$Z2$ sets of vector-like particles as 
Type I and Type II flipped
$SU(5)\times U(1)_X$ models, respectively.
Although we focus in this paper on Type II model,
results for proton decay are not found to 
differ significantly between the Type I and Type II 
models. 

To give the TeV-scale masses to the vector-like particles,
we must forbid the GUT scale or string scale masses for
the vector-like particles by some additional symmetries.
There are two solutions for this problem. In the first
solution, similar
to the next to the minimal supersymmetric SM (NMSSM),
we introduce a SM singlet Higgs field $S$ and a discrete
$Z_3$ symmetry. Thus, the heavy mass terms for these 
vector-like particles are forbidden by the $Z_3$ symmetry. 
Also, we consider the following superpotential
\begin{eqnarray}
W &=& \lambda'_3 S {\overline{XF}} XF 
+ \lambda'_4 S {\overline{Xl}} Xl ~.~\,
\end{eqnarray}
After $S$ acquires a vacuum expectation value (VEV) around
the TeV scale, these vector-like particles obtain the
TeV-scale masses. In the second solution, we can use
the Giudice-Masiero mechanism~\cite{Giudice:1988yz}. 
In the F-theory model
building, the discussions on the vector-like particle
masses  are similar to those
 on $\mu$ problem in Ref.~\cite{Heckman:2008qt}. 
We emphasize that
we might need to put the vector-like particles
$XF$ and ${\overline{XF}}$
 on different matter curves, and put $Xl$ and ${\overline{Xl}}$
 on different matter curves in F-theory model building.



{\bf Proton Decay~--}~Let us first review the
existing and proposed proton decay experiments.
Super-Kamiokande, a 50-kiloton (kt) water Cherenkov detector,
has set the current 
lower bounds of $8.2\times 10^{33}$ and $6.6\times 10^{33}$ years
at the $90\%$ confidence level for the partial lifetimes in the
$p\rightarrow e^+ \pi^0$ and $p\rightarrow \mu^+ \pi^0$ modes~\cite{:2009gd}.
Hyper-Kamiokande is a proposed 1-Megaton detector, about 20 times larger
volumetrically than Super-Kamiokande~\cite{Nakamura:2003hk},
which we can expect to explore partial lifetimes 
up to a level near $2\times 10^{35}$ years for
$p\rightarrow e^+ \pi^0$ across a decade long run.
The proposal for the DUSEL experiment~\cite{DUSEL}
features both water Cherenkov and liquid Argon (which is around five times
more sensitive per kilogram to $p\rightarrow K^+ {\bar \nu}_{\mu}$ than water)
detectors, in the neighborhood of 500 and 100 kt respectively,
with the stated goal of probing partial lifetimes into the
order of $10^{35}$ years for both the $e^+ \pi^0$ 
and $K^+{\bar \nu}_{\mu}$ channels.

Let us now specifically discuss the proton decay mode
$p\rightarrow e^+ \pi^0$ in flipped $SU(5)\times U(1)_X$ models.
After integrating out the heavy gauge boson
fields, we obtain the effective dimension six operator for proton decay 
\begin{eqnarray}
{ {\cal L}} & =  & \frac{g_{23}^2\epsilon^{ijk}}{2 M_{32}^2}   
 \left[ (
({\bar d^c}_{k}  \cos \theta_c
 + {\bar s^c}_{k} \sin \theta_c)
\gamma^\mu P_L u_{j}) 
\right. \nonumber \\
 &\times & (u_{i}
 \gamma_\mu P_L e_L)
 \left. +  h.c. \right] 
\label{Eq-1}
\end{eqnarray}
where $g_{23}$ is the $SU(3)_C\times SU(2)_L$ unified gauge coupling,
$\theta_c$ is the Cabibbo angle, and $u$, $d$, $s$, and $e$ are the
up quark, down quark, strange quark and electron, respectively. 
Also,  we neglect irrelevant CP-violating phases.

The decay amplitude is proportional to the overall normalization of the 
proton wave function at the origin.
Relevant matrix elements have been calculated
in a lattice approach~\cite{Kuramashi:2000hw}
with quoted errors below 10\%, corresponding to an
uncertainty of less than 20\% in the proton partial lifetime,
negligible compared to other uncertainties present in our calculation.
From Eq.~(\ref{Eq-1}), the proton lifetime is seen to scale as a fourth power
of the $SU(5)$ unification scale $M_{23}$, and inversely, again in the fourth
power, to the coupling $g_{23}$ evaluated at that scale.
This extreme sensitivity argues for great care in the
selection and study of a unification scenario.




{\bf Numerical Results~--}~
We have significantly upgraded a prior analysis of gauge coupling 
unification~\cite{Ellis:2002vk},
correcting a subtle inconsistency in usage of the effective Weinberg angle,
improving resolution of the light threshold corrections,
and undertaking a proprietary determination of the second loop,
starting fresh from the standard renormalization group 
equations (RGEs), {\it cf.}~\cite{Jiang:2006hf}.
The step-wise entrance of the top quark and 
supersymmetric particles (supersymmetric
partners of the SM particles) into the RGE
running is now properly accounted to all three gauge couplings individually 
rather than to a single composite term for the effective shift.
The two-loop contribution is likewise individually numerically determined
for each gauge coupling, including the top and bottom quark Yukawa couplings, 
taken themselves in the first loop.  
All three gauge couplings are
integrated recursively with the second loop into the Yukawa renormalization, with
the boundary conditions at the $Z$ boson mass $M_{Z}$ treated correctly for 
various values of $\tan \beta$, the ratio of Higgs vacuum expectation values.
The light threshold correction terms are included wherever the gauge 
couplings $\alpha_i$ are used.
Recognizing that the second loop itself influences the upper limit $M_{23}$
of its own integrated contribution, this feedback is accounted for in the dynamic
calculation of the unification scale~\cite{LNW-LP}.

In addition to the light $M_Z$-scale threshold corrections from the 
supersymmetric particles'
 entry into the RGEs, there may also be shifts occuring 
near the $M_{23}$ scale due to the heavy triplet Higgs fields and 
heavy gauge fields of $SU(5)$.
The light fields carry strong correlations to cosmology and low energy
phenomenology, so that we are guided toward plausible estimates of 
their mass distribution. For simplicity, we  consider the benchmark 
scenarios proposed in Ref.~\cite{Battaglia:2003ab}, which 
respect all available experimental constraints.
The heavy threshold corrections from the heavy triplet Higgs fields and 
heavy gauge fields, which can be quite substantial, are
much more difficult to constrain. 
Invoking naturalness, we assume 
\begin{eqnarray}
{{\sqrt {\lambda_1 \lambda_2}}\over 3} \le g_{23} \le 
3 {\sqrt {\lambda_1 \lambda_2}}~.~\, \label{eq:natural}
\end{eqnarray}
Moreover, the vector-like particles $XF$ and $\overline{XF}$
form complete $SU(5)\times U(1)_X$ multiplets, and the
contributions to the RGE running for the $SU(2)_L$ and $SU(3)_C$
gauge couplings from the vector-like particles $Xl$ and $\overline{Xl}$
are negligible. Thus, we assume degeneracy of these vector-like particles'
masses at a central value of 1 TeV.



\begin{figure}[htb]
\centering
\includegraphics[width=8.4cm]{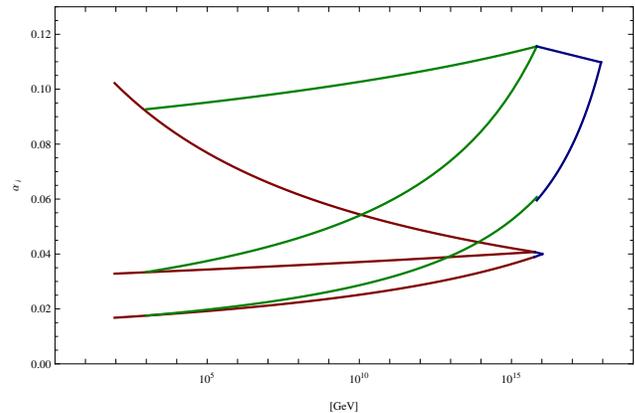}
\caption{Gauge coupling unification in the
minimal  (red solid lines) and  Type II  
(green solid lines) flipped $SU(5)\times U(1)_X$ models 
for benchmark scenario $B'$.
Starting from the top, we depict the gauge couplings $\alpha_3$, 
$\alpha_2$, and $\alpha_Y$.
The discontinuity at $M_Z$ (most visible for $\alpha_3$) stems from early
absorption of the thresholds into a function which is from that 
scale upward continuous.}
\label{fig1:GCU}
\end{figure}



In our numerical calculations, we use the weak-scale data in  
Ref.~\cite{Amsler:2008zzb}, and the top quark mass in 
Ref.~\cite{:2009ec}.
We adopt benchmark scenario $B'$ of Ref.~\cite{Battaglia:2003ab}
as our reference supersymmetric spectrum,
which is near a region of parameter space favored by the $\chi^2$ minimization
of cumulative deviation from experiments~\cite{Ellis:2004tc}.
We present gauge coupling unification for the minimal and
Type II flipped $SU(5)\times U(1)_X$ models
in Fig.~\ref{fig1:GCU}.
We additionally present 
the $U(1)_X$ gauge coupling $g_1$ at $M_{23}$,
unified $SU(5)$ coupling $g_{23}$,
mass scale $M_{23}$, and the proton partial lifetime
for the minimal, Type I and Type II models in Table~\ref{RGE-Data}.
Because of the TeV-scale vector-like particles,
we find parity for the gauge couplings $g_{23}$ in the Type I and Type II 
models, with each coupled significantly more strongly than the minimal model,
while $M_{23}$ is slightly larger. Thus, the proton partial
lifetime in the Type I and Type II models are shorter than
the minimal model by a factor 1/4.3.
The central prediction of the proton partial lifetime for the minimal, Type I 
and Type II models is
well below $10^{35}$ years, within the
reach of the future Hyper-Kamiokande and DUSEL experiments. However,
the uncertainty from heavy threshold corrections  
ever threatens to undo this promising result.



\begin{table}[htb]
\begin{center}
\begin{tabular}{|c|c|c|c|c|}
\hline
Model  & $g_1$ & $g_{23}$ & 
 $M_{23}$ (GeV)  &  $\tau_p$ (Years) \\
\hline
Minimal  & 0.70 & 0.72 & $5.8\times 10^{15}$  & $4.3\times 10^{34}$ \\
Type I & 0.75 & 1.21 & $6.8\times 10^{15}$ & $1.0\times 10^{34}$ \\
Type II & 0.87 & 1.20 & $6.8\times 10^{15}$ & $1.0\times 10^{34}$ \\
\hline
\end{tabular}
\end{center}
\caption{Gauge couplings $g_1$ and $g_{23}$, 
mass scale $M_{23}$, and proton partial lifetime $\tau_p$
in the minimal, Type I and Type II flipped 
$SU(5)\times U(1)_X$ models for  benchmark scenario $B'$.}
\label{RGE-Data}
\end{table}



\begin{figure}[htb]
\centering
\includegraphics[width=8.4cm]{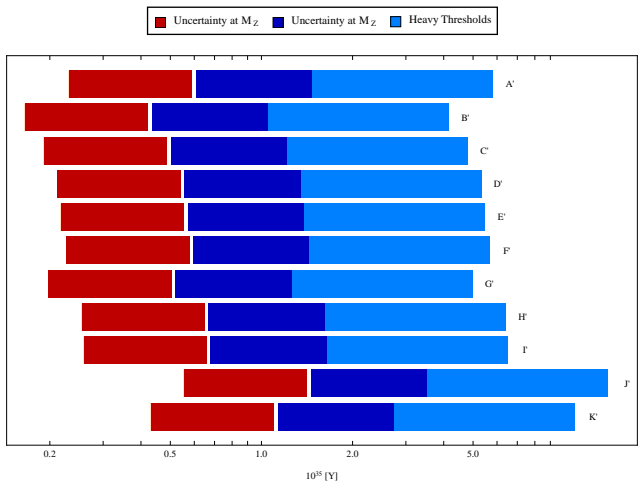}
\caption{Proton partial lifetime in the unit $10^{35}$ years
in the minimal flipped $SU(5)\times U(1)_X$ model.}
\label{fig2:FMSSM}
\end{figure}

\begin{figure}[htb]
\centering
\includegraphics[width=8.4cm]{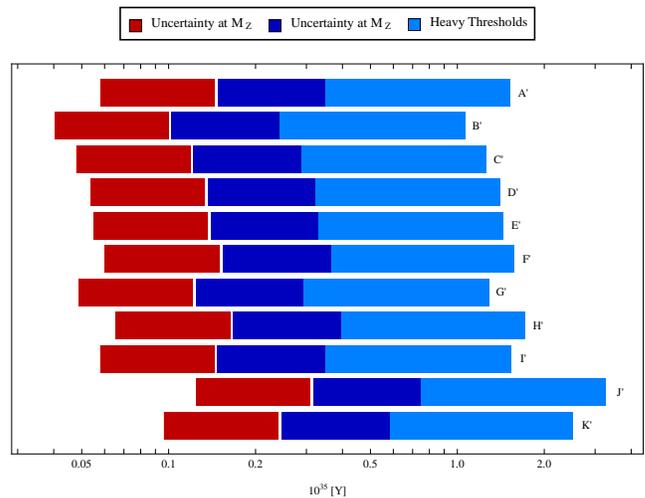}
\caption{Proton partial lifetime in the unit $10^{35}$ years
in the Type II flipped $SU(5)\times U(1)_X$ model.}
\label{fig3:FSU5}
\end{figure}

Including uncertainties from threshold corrections 
at the $M_Z$ and $M_{23}$ scales,
we present the proton partial lifetime in the minimal and
Type II flipped $SU(5)\times U(1)_X$ models for the process 
$p\rightarrow e^+ \pi^0$
in Figs.~\ref{fig2:FMSSM} and \ref{fig3:FSU5} respectively,
for each benchmark scenario from $A'$ to $K'$ 
of Ref.~\cite{Battaglia:2003ab}.
Central values are depicted by the narrow white gap between red and blue,
with the darkend regions on either side showing the error 
propagated from uncertainty in the $M_Z$-scale parameters,
combined in quadrature.
The lighter blue on the right-hand side depicts plausible
variation from the heavy threshold corrections, 
as in Eq.~(\ref{eq:natural}),
which can only extend the proton lifetime for
 flipped $SU(5)\times U(1)_X$ models.
In the minimal model, the central partial lifetime is in the range of 
$4-7\times 10^{34}$ years for benchmark scenarios from
$A'$ to $I'$, and about $1-2\times 10^{35}$ years for benchmark scenarios 
$J'$ and $K'$. However, the uncertainties from 
the heavy threshold corrections at $M_{23}$
are indeed quite large.
Proton decay appears to be within the reach of the future Hyper-Kamiokande
and DUSEL experiments if the heavy threshold
corrections are more modest.



For Type II flipped $SU(5)\times U(1)_X$ model, 
the central values for the partial lifetime are about 
$1-2\times 10^{34}$ years for benchmark scenarios 
from $A'$ to $I'$, and
about $2-3\times 10^{34}$ years for benchmark scenarios $J'$ and $K'$.
Even including uncertainties from the light and heavy threshold
corrections, the lifetime is still
less than  $2-3\times 10^{35}$ years for all scenarios considered.
A strong majority of the parameter space
for proton decay does indeed appear to be within the reach of 
the future Hyper-Kamiokande and DUSEL experiments for
the Type II flipped $SU(5)\times U(1)_X$ model.
This basic conclusion holds also for the
Type I flipped $SU(5)\times U(1)_X$ model.



{\bf Conclusions~--}~Proton decay is one of the most unique
yet ubiquitous predictions of GUTs.
We have studied the proton decay process 
$p\rightarrow e^+\pi^0$ via dimension six operators from the 
heavy gauge boson exchange.
Including uncertainties from the light and heavy
threshold corrections, we have shown that a majority of the
parameter space for proton decay is indeed within the reach of 
the future  Hyper-Kamiokande and DUSEL experiments for the
Type I and Type II flipped $SU(5)\times U(1)_X$ models.
The minimal flipped $SU(5)\times U(1)_X$ model is also
testable if the heavy threshold corrections are small.
In particular, detectability of TeV-scale vector-like particles
at the LHC presents an opportunity for cross correlation of results
between the most exciting particle physics experiments of the coming decade.

{\bf Acknowledgments~--}~We would like to thank D. B. Cline for
helpful private communication. This research was supported in part 
by  the DOE grant DE-FG03-95-Er-40917 (TL and DVN),
by the Natural Science Foundation of China 
under grant No. 10821504 (TL),
and by the Mitchell-Heep Chair in High Energy Physics (TL).



\begin{thebibliography}{99}






\bibitem{Langacker:1991an}
  J.~R.~Ellis, S.~Kelley and D.~V.~Nanopoulos,
  Phys.\ Lett.\ B {\bf 260}, 131 (1991);
  P.~Langacker and M.~X.~Luo,
  Phys.\ Rev.\ D {\bf 44}, 817 (1991);
  U.~Amaldi, W.~de Boer and H.~Furstenau,
  Phys.\ Lett.\ B {\bf 260}, 447 (1991).




\bibitem{:2009gd}
  H.~Nishino {\it et al.}  [Super-Kamiokande Collaboration],
  Phys.\ Rev.\ Lett.\  {\bf 102}, 141801 (2009).


\bibitem{Georgi:1974sy}
  H.~Georgi and S.~L.~Glashow,
  Phys.\ Rev.\ Lett.\  {\bf 32}, 438 (1974).

\bibitem{Dimopoulos:1981zb}
  S.~Dimopoulos and H.~Georgi,
  Nucl.\ Phys.\  B {\bf 193}, 150 (1981).


\bibitem{Murayama:2001ur}
  H.~Murayama and A.~Pierce,
  Phys.\ Rev.\  D {\bf 65}, 055009 (2002),
  and references therein.



\bibitem{Acharya:2001gy}
  B.~S.~Acharya and E.~Witten,
  arXiv:hep-th/0109152.

\bibitem{Witten:2001bf}
  E.~Witten,
  arXiv:hep-ph/0201018.


\bibitem{Beasley:2008dc}
  C.~Beasley, J.~J.~Heckman and C.~Vafa,
  JHEP {\bf 0901}, 058 (2009);
  JHEP {\bf 0901}, 059 (2009).


\bibitem{Donagi:2008ca}
  R.~Donagi and M.~Wijnholt,
  arXiv:0802.2969 [hep-th];
  arXiv:0808.2223 [hep-th].



\bibitem{smbarr} S. M. Barr,
Phys.\ Lett.\ B {\bf 112}, 219 (1982).


\bibitem{dimitri}
J.~P.~Derendinger, J.~E.~Kim and D.~V.~Nanopoulos,
Phys.\ Lett.\ B {\bf 139}, 170 (1984).

\bibitem{AEHN-0}
I.~Antoniadis, J.~R.~Ellis, J.~S.~Hagelin and D.~V.~Nanopoulos,
Phys.\ Lett.\ B {\bf 194}, 231 (1987).

\bibitem{Ellis:1995at}
  J.~R.~Ellis, J.~L.~Lopez and D.~V.~Nanopoulos,
  Phys.\ Lett.\  B {\bf 371}, 65 (1996).

\bibitem{Ellis:2002vk}
  J.~R.~Ellis, D.~V.~Nanopoulos and J.~Walker,
  Phys.\ Lett.\  B {\bf 550}, 99 (2002).

\bibitem{Jiang:2006hf}
  J.~Jiang, T.~Li and D.~V.~Nanopoulos,
  Nucl.\ Phys.\  B {\bf 772}, 49 (2007).

\bibitem{Lopez:1992kg}
  J.~L.~Lopez, D.~V.~Nanopoulos and K.~J.~Yuan,
  Nucl.\ Phys.\  B {\bf 399}, 654 (1993).


\bibitem{Jiang:2008yf}
  J.~Jiang, T.~Li, D.~V.~Nanopoulos and D.~Xie,
  Phys.\ Lett.\  B {\bf 677}, 322 (2009);
  Nucl.\ Phys.\  B {\bf 830}, 195 (2010).



\bibitem{Nakamura:2003hk}
  K.~Nakamura,
  Int.\ J.\ Mod.\ Phys.\  A {\bf 18}, 4053 (2003).

\bibitem{DUSEL}
  S.~Raby {\it et al.},
  arXiv:0810.4551 [hep-ph].

\bibitem{LNW-LP} 
T.~Li, D.~V.~Nanopoulos and J.~W.~Walker,
  arXiv:1003.2570 [hep-ph].


\bibitem{Giudice:1988yz}
  G.~F.~Giudice and A.~Masiero,
  Phys.\ Lett.\  B {\bf 206}, 480 (1988).


\bibitem{Heckman:2008qt}
  J.~J.~Heckman and C.~Vafa,
  JHEP {\bf 0909}, 079 (2009).


\bibitem{Kuramashi:2000hw}
  Y.~Kuramashi  [JLQCD Collaboration],
  arXiv:hep-ph/0103264.

\bibitem{Battaglia:2003ab}
  M.~Battaglia, A.~De Roeck, J.~R.~Ellis, F.~Gianotti, K.~A.~Olive and L.~Pape,
  Eur.\ Phys.\ J.\  C {\bf 33}, 273 (2004).



\bibitem{Amsler:2008zzb}
  C.~Amsler {\it et al.}  [Particle Data Group],
  Phys.\ Lett.\  B {\bf 667}, 1 (2008).


\bibitem{:2009ec}
    [Tevatron Electroweak Working Group and CDF Collaboration and D0 Collab],
  arXiv:0903.2503 [hep-ex].

\bibitem{Ellis:2004tc}
  J.~R.~Ellis, S.~Heinemeyer, K.~A.~Olive and G.~Weiglein,
  JHEP {\bf 0502}, 013 (2005).





\end{thebibliography}
\end{document}